\documentclass{article}

\usepackage{arxiv}

\usepackage[utf8]{inputenc} % allow utf-8 input
\usepackage[T1]{fontenc}    % use 8-bit T1 fonts
\usepackage{hyperref}       % hyperlinks
\usepackage{url}            % simple URL typesetting
\usepackage{booktabs}       % professional-quality tables
\usepackage{amsfonts}       % blackboard math symbols
\usepackage{nicefrac}       % compact symbols for 1/2, etc.
\usepackage{microtype}      % microtypography
\usepackage{lipsum}		% Can be removed after putting your text content
\usepackage{graphicx}
\usepackage[round]{natbib}
\usepackage{doi}

\usepackage{mathtools,upgreek,eucal,mathrsfs,enumitem,amsthm,amsmath}
\usepackage{times,epsfig,makeidx,amsfonts,amsmath,dcolumn,multirow,amssymb,colortbl,bm,url,color}
\usepackage{algorithm}
\usepackage[]{algorithmic}
 \usepackage[flushleft]{threeparttable}
\usepackage{subfig}
\usepackage{mathrsfs}

\newtheorem{definition}{Definition}

\newcommand{\bxi}{\bm{\bxi}}

\newcommand{\btheta}{\bm{\theta}}

\newcommand{\bo}{{\mathcal O}}

\newtheorem{proposition}{Proposition}

\title{On harmonic oscillator hazard functions}
\author{ 
	\href{https://orcid.org/0000-0002-5795-4345}{\includegraphics[scale=0.06]{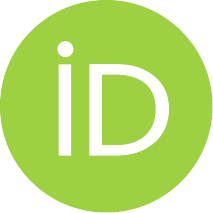}\hspace{1mm}J. Andr{\' e}s Christen} \\
	Department of Probability and Statistics\\
	Centro de Investigaci{\' o}n en Matem{\' a}ticas (CIMAT-CONAHCYT) \\
	Guanajuato, M{\' e}xico\\
	\texttt{jac@cimat.mx} 
\And	
	\href{https://orcid.org/0000-0001-7183-8407}{\includegraphics[scale=0.06]{orcid.pdf}\hspace{1mm}Francisco Javier Rubio} \\
	Department of Statistical Science\\
	University College London \\
	London, UK\\
	\texttt{f.j.rubio@ucl.ac.uk} 
	}
\hypersetup{
pdftitle={XXX},
pdfsubject={stat.ME, stat.AP},
pdfauthor={Christen, Rubio},
}

\usepackage{color}
\definecolor{rojo}{RGB}{255,0,0}
\definecolor{azul}{RGB}{0,0,255}
\definecolor{verde}{RGB}{0,135,0}
\newcommand{\rev}[1]{\textcolor{red}{\textbf{#1}}}
\begin{document}
\maketitle

\begin{abstract}
We propose a parametric hazard model obtained by enforcing positivity in the damped harmonic oscillator. The resulting model has closed-form hazard and cumulative hazard functions, facilitating likelihood and Bayesian inference on the parameters. We show that this model can capture a range of hazard shapes, such as increasing, decreasing, unimodal, bathtub, and oscillatory patterns, and characterize the tails of the corresponding survival function. We illustrate the use of this model in survival analysis using real data. 
\end{abstract}

% keywords can be removed

%----------------------------------------------------------------------------------------------------------------------------------------------------------------------

%%%%%%%%%%%%%%%%%%%%%%%%%%%%%%%%%%%%%%%%%%%%%%%%%%%%%%%%%%%%%%%%%%%%%%%%%%%%%%%%%%%%%%%%%%%%%%%%%%%%%%
%%%%%%%%%%%%%%%%%%%%%%%%%%%%%%%%%%%%%%%%%%%%%%%%%%%%%%%%%%%%%%%%%%%%%%%%%%%%%%%%%%%%%%%%%%%%%%%%%%%%%%
\section{Introduction}\label{sec:intro}
%%%%%%%%%%%%%%%%%%%%%%%%%%%%%%%%%%%%%%%%%%%%%%%%%%%%%%%%%%%%%%%%%%%%%%%%%%%%%%%%%%%%%%%%%%%%%%%%%%%%%%
%%%%%%%%%%%%%%%%%%%%%%%%%%%%%%%%%%%%%%%%%%%%%%%%%%%%%%%%%%%%%%%%%%%%%%%%%%%%%%%%%%%%%%%%%%%%%%%%%%%%%%

The hazard function plays a key role in the analysis of survival data \citep{rinne:2014}. {For a positive random variable $T$, with probability density function $f(t)$ and cumulative distribution $F(t)$, the hazard function is defined as:
\begin{equation}
h(t) = \lim_{\Delta t \downarrow 0} \dfrac{{\mathbb P}\left(t \leq T < t + \Delta t \mid T \geq t\right)}{\Delta t} = \dfrac{f(t)}{1-F(t)}.
\label{eq:hazard_function}
\end{equation}
}
Given the intuitive interpretation of the hazard function as the instantaneous failure rate at time $t>0$, this function serves as basis for defining many survival regression models \cite{rubio:2019}. Estimation of the hazard function using both parametric and non-parametric methods has received considerable attention. We refer the reader to \cite{christen:2024} for a comprehensive overview of these methods, which include the use of parametric distributions, splines, and Bayesian nonparametric methods. In the parametric setting, it is desirable to define models capable of capturing the basic shapes of the hazard function: increasing, decreasing, unimodal (up-then-down), and bathtub (down-then-up). To this end, several generalizations of the Weibull distribution have been proposed, including the generalized gamma distribution, the exponentiated Weibull distribution, and the power generalized Weibull distribution, among others. Similarly, general methods have been proposed to add parameters to a baseline distribution or baseline hazard function. See \cite{marshall:1997} and \cite{anaya:2021} for an overview of these methods. More recently, \cite{christen:2024} proposed parametrically modeling the hazard function using a system of first-order autonomous ordinary differential equations (ODEs) with positive solutions. This approach offers a general methodology for generating distributions with interpretable parameters and allows for adding flexibility to the resulting solutions by including hidden states.

Following \cite{christen:2024}, a novel family of hazard functions is derived using a general linear second-order differential equation that represents the classical damped harmonic oscillator.
An additional parameter is used to allow the system to reach the equilibrium point at a positive value (in contrast to the classical damped harmonic oscillator, which stabilizes at zero). The resulting model, which can be reduced to three parameters by fixing the initial conditions (as discussed later), contains parameters that influence the shape of the hazard function, resulting in a flexible, tractable, parsimonious (three parameters only) and interpretable model. We characterize the tails of the survival function and the hazard shapes in terms of the parameter values. 
We present a real data example which shows how the model can capture complex hazard shapes, and that it provides a better fit compared to appropriate competitor models. R code and data are provided in the GitHub repository \url{https://github.com/FJRubio67/HOH}. Python code may also be obtained from JAC. All proofs are presented in the appendix.

%%%%%%%%%%%%%%%%%%%%%%%%%%%%%%%%%%%%%%%%%%%%%%%%%%%%%%%%%%%%%%%%%%%%%%%%%%%%%%%%%%%%%%%%%%%%%%%%%%%%%%
%%%%%%%%%%%%%%%%%%%%%%%%%%%%%%%%%%%%%%%%%%%%%%%%%%%%%%%%%%%%%%%%%%%%%%%%%%%%%%%%%%%%%%%%%%%%%%%%%%%%%%
\section{Modelling the hazard function with a harmonic oscillator}\label{sec:model}
%%%%%%%%%%%%%%%%%%%%%%%%%%%%%%%%%%%%%%%%%%%%%%%%%%%%%%%%%%%%%%%%%%%%%%%%%%%%%%%%%%%%%%%%%%%%%%%%%%%%%%
%%%%%%%%%%%%%%%%%%%%%%%%%%%%%%%%%%%%%%%%%%%%%%%%%%%%%%%%%%%%%%%%%%%%%%%%%%%%%%%%%%%%%%%%%%%%%%%%%%%%%%
Let us first introduce some notation and assumptions. Let ${\bf o} = (o_1,\dots,o_n)$ be a sequence of survival times, $c_i \in \mathbb{R}_+$ the right-censoring times, $t_i=\min\{o_i,c_i\}$ be the observed times, and $\delta_i=\mbox{I}(o_i\leq c_i)$ be the indicator that observation $i$ is uncensored, $i=1,\dots,n$.
Suppose that the survival times are generated by a continuous probability distribution, with twice continuously differentiable probability density function $f(t)$, cumulative distribution function $F(t) = \int_0^t f(r) dr$, survival function $S(t) = 1-F(t)$, hazard function $h(t) = \frac{f(t)}{S(t)}$, and cumulative hazard function $H(t) = \int_0^t h(r) dr$. 

{Based on definition \eqref{eq:hazard_function}, a hazard function for an absolutely continuous, positive, random variable may be defined through a continuous function satisfying $h(t) \geq 0$, and $\int_0^t h(r) dr < \infty$, for $t\geq 0$. Thus, any function satisfying these properties can be used as a hazard function. This allows for defining various types of hazard estimators or hazard-based regression models, such as those that model the hazard function using splines. Building on this idea}, we propose modelling the hazard function, $h(t)$, through the linear second order ODE:
\begin{equation}
h''(t) + 2 \eta w_0 h'(t) + w^2_0 (h(t) - h_b) = 0; \quad h(0) = h_0, \quad h'(0) = r_0.
\label{eq:harmonic_osc}    
\end{equation}
{In any case, once $h$ is postulated as a hazard function, this fixes the distribution of the survival times, \textit{i.e.}~the probability model involved, and should always be interpreted as in \eqref{eq:hazard_function}.}
Equation \eqref{eq:harmonic_osc} models the acceleration of the hazard function and also has a geometric interpretation, describing the curvature of the hazard function, both at any time $t > 0$. More specifically, this equation can be seen as a shifted version of the \textit{damped harmonic oscillator} \citep{georgi:1993,strogatz:2018}. The damped harmonic oscillator is a classical model in Physics which is used to describe the motion of a mass attached to a spring when damping (friction) is present \citep{georgi:1993,strogatz:2018}. The damped harmonic oscillator is generally understood as a system that degrades or stabilizes over time due to the combined effects of the restoring force and damping. Thus, the damped harmonic oscillator \eqref{eq:harmonic_osc} represents the evolution of the instantaneous failure rate, depicting a system that evolves over time to reach a degraded or stable state.
Since the solutions to the damped harmonic oscillator equation can take negative values, and the equilibrium point is exactly $0$, we consider the shifted version \eqref{eq:harmonic_osc} of this model to allow for a positive equilibrium point $h_b>0$. The parameters of this ODE are easily interpreted. The natural frequency, $w_0 > 0$, controls the frequency of oscillation, while the damping coefficient, $\eta > 0$, represents the dissipative forces acting on the system. The shift parameter, $h_b > 0$, represents the stability state or equilibrium level of the solution.
In the above parametrization the value of the damping coefficient $\eta$ determines the system's behavior and, in our case, the shape of the hazard function. The three regimens of the system are the under-damped case ($0 \leq \eta < 1$), the over-damped case ($\eta > 1$), and the critically damped case ($\eta = 1$). We present a more detailed analysis of these cases in the next Sections.

The ODE in \eqref{eq:harmonic_osc} defines a parametric hazard function $h(t \mid \btheta)$ with parameters $\btheta = (\eta, w_0, h_b, h_0, r_0)^{\top} \in {\mathbb R}_+^4 \times {\mathbb R}$. There are parameter values $\btheta$ that lead to a negative $h(t\mid \btheta)$ for some values of $t$. The reason for this is that the shift parameter $h_b>0$ only defines the equilibrium position of the hazard function, which may not be sufficient to translate the entire oscillator to the positive quadrant. In such cases, it is necessary to discard parameter values that lead to negative solutions. This approach is related to a method known as \textit{enforcing positivity} of ODE solutions, which is used in several areas \citep{shampine:2005, blanes:2022}. The parameter values leading to positive hazard functions will be referred to as the admissible parameter space, as follows.
\begin{definition}
Let $\btheta = (\eta, w_0, h_b, h_0, r_0)^{\top} \in {\mathbb R}_+^4 \times {\mathbb R}$. We define the \textit{admissible parameter space} as:
\begin{equation*}
\Theta_A = \{\btheta  \in {\mathbb R}_+^4 \times {\mathbb R}: h(t \mid \btheta) > 0 \quad \text{for all} \quad t \geq 0\}.    
\end{equation*}
\end{definition}
In the following Sections, we will present a simple characterization of the admissible parameter space $\Theta_A$ based on the parameter values.

%---------------------------------------------------------------------------------------------------------------------------------------------------
\subsection{Under-damped case}
%---------------------------------------------------------------------------------------------------------------------------------------------------
The analytic solution to \eqref{eq:harmonic_osc} in the under-damped case ($0 \leq \eta < 1$) is:
\begin{equation*}
h(t \mid \btheta) = h_b + A e^{- w_0 \eta t} \sin \left( w_1 t + \phi \right),
\end{equation*}
where $w_1 = w_0 \sqrt{| \eta^2 -1 |}$, $A$ denotes the amplitude of the oscillations, and $\phi$ is the phase, which represents the position of the hazard function within its cycle of oscillation. $h(t \mid \btheta) - h_b$ tends exponentially to zero as $t\to \infty$, since the term $e^{- w_0 \eta t}$ dominates the asymptotic behavior of $h(t \mid \btheta)$. 
From the initial conditions $h(0) = h_0$ and $h'(0) = r_0$ one can easily find the value for the amplitude $A$ and the phase $\phi$.  
That is, if $h_b \neq h_0$ then, if $r_0 + w_0 \eta ( h_0 - h_b) \neq 0$, $A = \dfrac{h_0 - h_b}{\sin(\phi)}$ and $ \phi = \tan^{-1} \left( \frac{w_1 (h_0 - h_b)}{r_0 + w_0 \eta ( h_0 - h_b)}  \right)$ and, if  $r_0 + w_0 \eta ( h_0 - h_b) = 0$, $\phi = \operatorname{sign}(h_0 - h_b) \frac{\pi}{2}$ and $A = h_0 - h_b$. If $ h_0 - h_b = 0$, then $\phi = 0$ and $A = \dfrac{r_0}{w_1}$. The cumulative hazard function can be found by integrating $h(t \mid \btheta)$ as follows:
\begin{equation*}
H(t \mid \btheta) = h_b t + \frac{A}{w_0 \eta} 
\frac{\sin(\phi) + \mu \cos(\phi) - e^{-  w_0 \eta t} (\sin(w_1 t + \phi) + \mu \cos(w_1 t + \phi))}{\mu^2 + 1} ,
\end{equation*}
where  $\mu = \frac{w_1}{w_0 \eta} = \frac{\sqrt{\vert \eta^2 - 1 \vert}}{\eta}$.

If $r_0 < 0$ and/or $h_0 > 2 h_b$, $h(t \mid \btheta)$ may become negative for a region around $0$, say $t \in [0,t_M]$. To test if is a set of parameters leads to $h(t \mid \btheta) < 0$ for $t \in [0,t_M]$, we can calculate the derivative of the analytic solution to obtain
\begin{equation*}
h'(t \mid \btheta) = -A e^{- w_0 \eta t} \left(  w_1 \cos \left( w_1 t + \phi \right) -w_0 \eta \sin \left( w_1 t + \phi \right) \right). 
\end{equation*}
Solving $h'(t \mid \btheta)=0$ leads to $\tan (w_1 t^* + \phi) = \mu$.  This equation has many solutions, but we know that the minimum must be in the first or second critical point after $t=0$, given that the solution has an envelope function $h_b \pm A e^{-w_0 \eta t}$. Therefore, we test these two critical points 
\begin{equation*}
t^* = \frac{\arctan(\mu) - \phi + (0 ~~\text{or}~~ \pi)}{w_1} ,
\end{equation*}
to find the minimum $h(t^* \mid \btheta)$. If $h(t^* \mid \btheta) < 0$, for either solution, then $\btheta \notin \Theta_A$, since this implies that $h(t \mid \btheta) < 0$ for some region $t \in [0,t_M]$. Otherwise, it follows that $h(t \mid \btheta) > 0$ for $t>0$ and $\btheta \in \Theta_A$.

%---------------------------------------------------------------------------------------------------------------------------------------------------
\subsection{Over-damped case}
%---------------------------------------------------------------------------------------------------------------------------------------------------

The case when $\eta > 1$ is called over-damped, since the state variable (hazard function) $h(t \mid \btheta)$ does not oscillate and returns, exponentially, to $h_b$.  
%There are two case $\eta = 1$, which is called ``critically damped'' and $\eta > 1$ the over damped, although the former has the same fundamental property of the latter, that $h$ returns exponentially to $h_b$.  We study first $\eta > 1$.
The analytic solution in this case is
\begin{equation*}
h(t \mid \btheta) = h_b + e^{-w_0 \eta t} \left(
\frac{h_0-h_b + a}{2} e^{ w_1 t} +
\frac{h_0-h_b - a}{2} e^{-w_1 t} \right),
\end{equation*}
where $a = \frac{h_0-h_b}{\mu} + \frac{r_0}{w_1} $ ($w_1 = w_0 \sqrt{ \eta^2 -1 }$ as above). Since $\sqrt{\eta^2 - 1} < \eta$ then $-w_0 \eta \pm w_1 < 0$ and $h(t \mid \btheta) - h_b$ also tends exponentially to zero, as $e^{-w_0 (\eta - w_1) t} $.
The cumulative hazard function is: 
\begin{equation*}
H(t \mid \btheta) = h_b t +  
\frac{h_0-h_b + a}{2 (-w_0 \eta + w_1)}  (e^{ (-w_0 \eta + w_1) t} - 1) +
\frac{h_0-h_b - a}{2 (-w_0 \eta - w_1)} (e^{(-w_0 \eta - w_1) t} -1) .
\end{equation*}
Depending on the value of $r_0$, $h(t \mid \btheta)$ may take negative values for a region $t\in [0,t_M]$. In this case, we know there is at most one critical point, as the solution does not oscillate \citep{georgi:1993,strogatz:2018}. By solving $h'(t \mid \btheta)=0$, the critical point is
\begin{equation*}
t^* = \frac{1}{2 w_1}
\log \left( \frac{(h_0-h_b -a)(w_1+w_0\eta)}{(h_0-h_b +a)(w_1-w_0\eta)}  \right),
\end{equation*}
if $\frac{(h_0-h_b -a)}{(h_0-h_b +a)(w_1-w_0\eta)} > 0$, otherwise there is no critical point. Moreover, if $r_0>0$, the solution is increasing or unimodal (see Proposition \ref{prop:shapes}), which implies that $h(t \mid \btheta) > 0$ for $t>0$, and then $\btheta \in \Theta_A$. If $r_0 < 0$, depending on the values of the remaining parameters, the shape of the hazard function can be decreasing or bathtub (see Proposition \ref{prop:shapes}). We can characterize the admissible parameter space as follows. If $\frac{(h_0-h_b -a)}{(h_0-h_b +a)(w_1-w_0\eta)} > 0$ and $h(t^* \mid \btheta) <0$, then $\btheta \notin \Theta_A$. If $\frac{(h_0-h_b -a)}{(h_0-h_b +a)(w_1-w_0\eta)} > 0$ and $h(t^* \mid \btheta) >0$, then $\btheta \in \Theta_A$. If $\frac{(h_0-h_b -a)}{(h_0-h_b +a)(w_1-w_0\eta)} < 0$, then $\btheta \in \Theta_A$.

%---------------------------------------------------------------------------------------------------------------------------------------------------
\subsection{Critically-damped case}
%---------------------------------------------------------------------------------------------------------------------------------------------------
The case $\eta = 1$ is referred to as critically damped. The analytical solution is
\begin{equation*}
h(t \mid \btheta) = h_b + \left\{ h_0-h_b + t\left( r_0 + w_0 (h_0-h_b) \right) \right\} e^{- w_0  t } .
\end{equation*}
% \begin{equation*}
% h(t) = h_b + \left\{ h_0-h_b + t\left( r_0 + w_0 \eta (h_0-h_b) \right) \right\} e^{- w_0 \eta t } .
% \end{equation*}
The corresponding cumulative hazard is:
\begin{equation*}
H(t\mid \btheta) = h_b t + \dfrac{(h_0-h_b) \left(1 - e^{- w_0 t } \right)}{  w_0} + 
\dfrac{\left( r_0 + w_0 (h_0-h_b) \right)}{w_0^2 } \left( e^{w_0 t} - w_0 t - 1 \right) e^{-w_0 t}.
\end{equation*}
% \begin{equation*}
% H(t) = h_b t + \dfrac{(h_0-h_b) \left(1 - e^{-\eta w_0 t } \right)}{\eta  w_0} + 
% \dfrac{\left( r_0 + w_0 \eta (h_0-h_b) \right)}{w_0^2 \eta^2} \left( e^{w_0 \eta t} - w_o \eta t - 1 \right) e^{-w_0 \eta t}
% \end{equation*}
Unfortunately, there is no closed-form expression for the critical point in this case. However, since this is a limit case of the under-damped and over-damped cases, and we will consider continuous prior distributions for $\eta$ in the Bayesian analysis presented in Section \ref{sec:example}, this case will have zero probability.  Consequently, it is ignored in our implementation. Nonetheless, the expressions for the hazard and cumulative hazard functions are presented here for completeness.

%---------------------------------------------------------------------------------------------------------------------------------------------------
\subsection{Parametrization}
%---------------------------------------------------------------------------------------------------------------------------------------------------

Overall, we adopt the parametrization \eqref{eq:harmonic_osc}, with parameters $\eta$, $w_0$, $h_b$, $h_0$, and $r_0$. Nonetheless, other parametrizations might provide further insights into the model. Looking at the ODE formulation in \ref{eq:harmonic_osc}, $w_0$ is called the ``natural frequency'' and its units are the inverse of time units.  The parameter $\eta$ has no units, and is referred to as the ``damping ratio''. Then, $w_1 = w_0 \sqrt{| \eta^2 -1 |}$ is the angular frequency in the under-damped case, and has the same units as $w_0$.  This means that if $t \in [ 0, t_M]$, $t_M>0$, $h$ will oscillate $\frac{w_1 t_M}{2 \pi}$ times in that time window.

Since the hazard function $h(t\mid \btheta)$ does not contain an explicit scale parameter, it is important to check if one can control the scale of this function through a combination of the parameters.  It can be noted that by using the following equivalent parametrization, $(h_b, \eta, w_0, \alpha_0, \theta_0)$ for \eqref{eq:harmonic_osc}
\begin{equation*}
h''(t) + 2 \eta (w_0 h_b) h'(t) + (w_0 h_b)^2 (h(t) - h_b) = 0, ~h(0)=\frac{\alpha_0}{h_b}, ~h'(0) = \frac{\theta_0}{h_b} ,
\end{equation*}
$h_b$ becomes a scale parameter in the usual sense.  Since both parametrizations are equivalent, the model does not need an additional scale parameter, and which parametrization to use is a matter of convenience.  In our applications, we will provide empirical strategies for fixing the initial conditions (see Section~\ref{sec:inference}) and therefore we use the original parametrization \eqref{eq:harmonic_osc}.

%%%%%%%%%%%%%%%%%%%%%%%%%%%%%%%%%%%%%%%%%%%%%%%%%%%%%%%%%%%%%%%%%%%%%%%%%%%%%%%%%%%%%%%%%%%%%%%%%%%%%%
%%%%%%%%%%%%%%%%%%%%%%%%%%%%%%%%%%%%%%%%%%%%%%%%%%%%%%%%%%%%%%%%%%%%%%%%%%%%%%%%%%%%%%%%%%%%%%%%%%%%%%
\section{Shapes and tail-weight characterization}\label{sec:weightshapes}
%%%%%%%%%%%%%%%%%%%%%%%%%%%%%%%%%%%%%%%%%%%%%%%%%%%%%%%%%%%%%%%%%%%%%%%%%%%%%%%%%%%%%%%%%%%%%%%%%%%%%%
%%%%%%%%%%%%%%%%%%%%%%%%%%%%%%%%%%%%%%%%%%%%%%%%%%%%%%%%%%%%%%%%%%%%%%%%%%%%%%%%%%%%%%%%%%%%%%%%%%%%%%

\begin{figure}
\centering
\begin{tabular}{c c}
\includegraphics[scale=0.4]{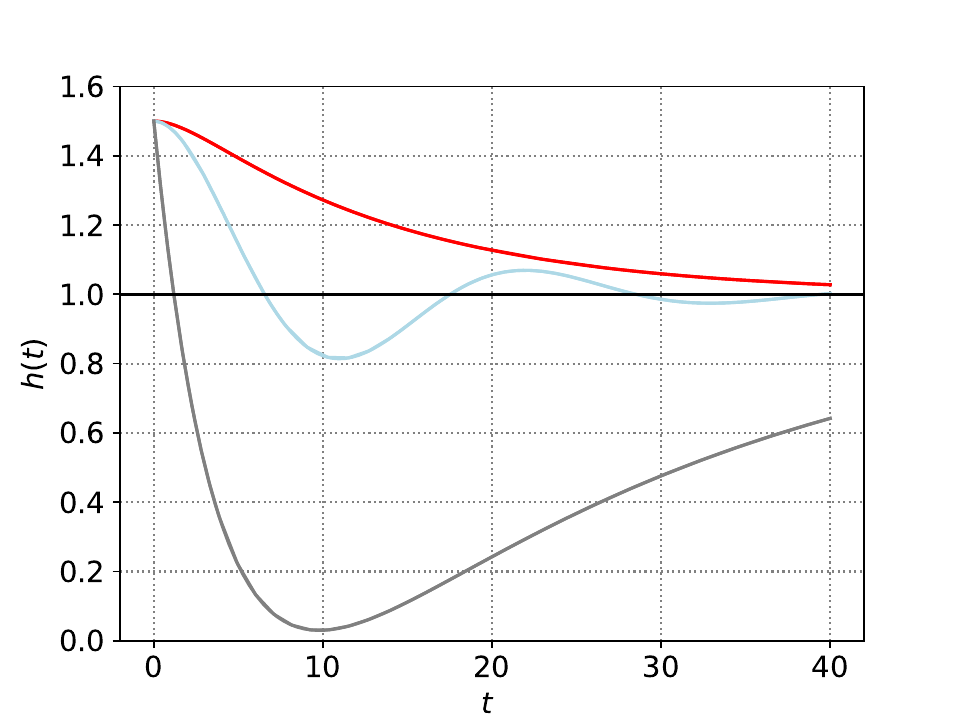} & \includegraphics[scale=0.4]{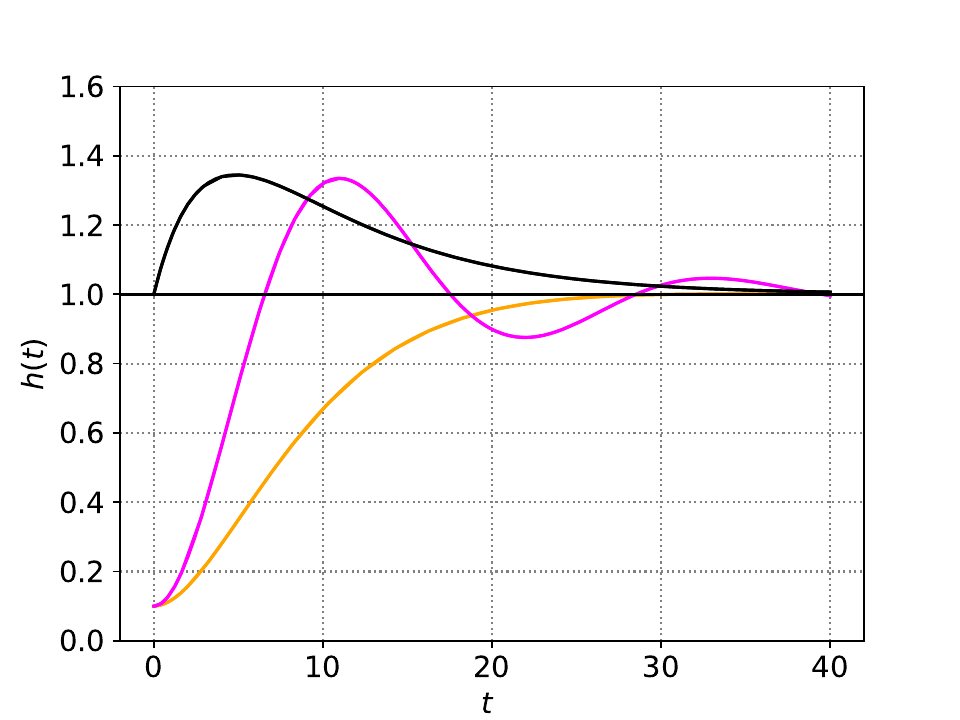} \\
 (a) & (b) \\
 \end{tabular}
\caption{
Examples of harmonic oscillator hazard functions: (a) Strictly decreasing (red, over-damped), ``bathtub'' (gray, over-damped) and oscillating (blue, under-damped). (b) Strictly increasing (orange, over-damped), unimodal (black, over-damped) and oscillating (magenta, under-damped).  All converge to the asymptotic value $h_b$ ($= 1$).
}
\label{fig:har_exa}
\end{figure}

As previously discussed, one of the desirable properties of parametric survival models is their ability to capture the basic shapes. The following result presents a characterization of the hazard shapes of \eqref{eq:harmonic_osc}, in terms of the parameter values, which include the basic shapes as well as oscillatory cases. We focus on the under-damped and over-damped cases, which are the main cases of interest for modelling, as discussed in the previous Section. See Figure \ref{fig:har_exa} for examples of the hazard shapes obtained as a solution to \eqref{eq:harmonic_osc}.

\begin{proposition}\label{prop:shapes}
Let $T>0$ be a random variable with hazard function defined by the damped harmonic oscillator model \eqref{eq:harmonic_osc}, and $\btheta \in \Theta_A$. Then, the corresponding hazard function can capture the following shapes:
\begin{itemize}
\item Increasing (monotonic): $\eta > 1$, $r_0 > 0$, $\frac{(h_0-h_b -a)}{(h_0-h_b +a)(w_1-w_0\eta)} < 0$.

\item Decreasing (monotonic): $\eta > 1$, $r_0 < 0$, $\frac{(h_0-h_b -a)}{(h_0-h_b +a)(w_1-w_0\eta)} < 0$.

\item Unimodal (up-then-down): $\eta > 1$, $r_0 > 0$, $\frac{(h_0-h_b -a)}{(h_0-h_b +a)(w_1-w_0\eta)} > 0$.

\item Bathtub (down-then-up): $\eta > 1$, $r_0 < 0$, $\frac{(h_0-h_b -a)}{(h_0-h_b +a)(w_1-w_0\eta)} > 0$.

\item Oscillatory (multiple cycles): $\eta<1$.
\end{itemize}
\end{proposition}

Although some parametric distributions, such as the power generalized Weibull or generalized gamma distributions, can capture the first four shapes, they do not account for oscillatory behaviors. {Another important distinction is that the hazard function of the damped harmonic oscillator is always upper-bounded, with its amplitude decaying over time. Consequently, $h(t)$ approaches the stable state $h_b$ as $t \to \infty$}. We acknowledge that these are implicit assumptions of the damped harmonic oscillator model, which may not be suitable for all types of data sets. However, we argue that all parametric models have implicit, and often overlooked, limitations. For instance, the Weibull distribution implicitly requires that when the hazard function is increasing, it must satisfy $h(0)=0$ and remain unbounded, which implies lighter tails than those of the exponential distribution \citep{christen:2024}. Conversely, decreasing Weibull hazard functions must satisfy $h(0)=\infty$. 
\rev{}
The proposed damped harmonic oscillator model allows for a better understanding of the hazard shapes and properties thanks to the interpretability of the parameters in the ODE \eqref{eq:harmonic_osc}.

The following proposition presents a characterization of the right tail of the distribution of a random variable with hazard function defined by the damped harmonic oscillator model \eqref{eq:harmonic_osc}. As we can see, the shifting parameter plays a key role in controlling the tail-weight in all cases. 

\begin{proposition}\label{prop:tails}
Let $T>0$ be a random variable with hazard function defined by the damped harmonic oscillator model \eqref{eq:harmonic_osc}, with $\btheta \in \Theta_A$. Then, 
\begin{equation*}
{\mathbb P}(T> t \mid \btheta)  = \bo \left(  e^{- h_b t } \right).
\end{equation*}
\end{proposition}
%\ac A proof of this proposition may be found in Appendix~\ref{sec:appendix}. \ca
The above result indicates that the distribution associated with the harmonic oscillator hazard function is a sub-exponential distribution \citep{vershynin:2018}. An important difference of model \eqref{eq:harmonic_osc} compared to the exponential distribution (which has a constant hazard function) or other parametric sub-exponential distributions \citep{vershynin:2018}, is that the proposed model \eqref{eq:harmonic_osc} allows for capturing a variety of shapes of the hazard function. This sub-exponential tail-weight characterization also allows for deriving other properties of the distribution induced by \eqref{eq:harmonic_osc}, such as properties of the moments and moment generating function, as detailed in Chapter 2 of \cite{vershynin:2018}, and thus omitted here.

%%%%%%%%%%%%%%%%%%%%%%%%%%%%%%%%%%%%%%%%%%%%%%%%%%%%%%%%%%%%%%%%%%%%%%%%%%%%%%%%%%%%%%%%%%%%%%%%%%%%%%
%%%%%%%%%%%%%%%%%%%%%%%%%%%%%%%%%%%%%%%%%%%%%%%%%%%%%%%%%%%%%%%%%%%%%%%%%%%%%%%%%%%%%%%%%%%%%%%%%%%%%%
\section{Inference}\label{sec:inference}
%%%%%%%%%%%%%%%%%%%%%%%%%%%%%%%%%%%%%%%%%%%%%%%%%%%%%%%%%%%%%%%%%%%%%%%%%%%%%%%%%%%%%%%%%%%%%%%%%%%%%%
%%%%%%%%%%%%%%%%%%%%%%%%%%%%%%%%%%%%%%%%%%%%%%%%%%%%%%%%%%%%%%%%%%%%%%%%%%%%%%%%%%%%%%%%%%%%%%%%%%%%%%
%\subsection{Likelihood}
The likelihood function for model \eqref{eq:harmonic_osc} under right-censoring is fully determined by the hazard and cumulative hazard functions, $h(t \mid \btheta)$ and $H(t \mid \btheta)$ as follows,
\begin{equation*}
\mathcal{L}(\btheta) =  \prod_{i=1}^n  h(t_i \mid \btheta)^{\delta_i} \exp\left\{ - H(t_i \mid \btheta) \right\}, \quad \btheta \in \Theta_A.
\end{equation*}
Since these functions are available in analytic form, the likelihood function can be implemented in any numeric programming language. This also implies that the maximum likelihood estimates can be calculated using general-purpose optimization methods, {although this may require comparing different methods an initial points since, in general, the concavity of the above likelihood (or log-likelihood) function is not guaranteed}.

%The initial conditions $h_0$ and $r_0$ can be assumed to be unknown parameters, or fixed using previous information about the phenomenon of interest.

In general, one could consider the initial conditions, $h_0$ and $r_0$, as unknown parameters to be estimated.
%, or they may be fixed from the onset based on prior information as follows. 
However, estimating these parameters may be challenging in practice for some data sets that do not contain uncensored observations near $t=0$. Alternatively, one can fix the initial conditions using prior or expert information (see \cite{christen:2024} for a discussion on these points).
Specifically, setting values for the survival function  $S(\Delta t)$ and $S(2\Delta t)$ for some (``small'') initial time step $\Delta t$,
we may approximate the initial condition $h_0= h(0)$ using 
\begin{equation*}
h_0  = -\dfrac{S'(0)}{S(0)} \approx -\dfrac{S'(\Delta t)}{S(\Delta t)} \approx  -\dfrac{S(\Delta t) - S(0)}{\Delta t S(\Delta t)}.
\end{equation*}
Similarly, we may approximate the initial condition $r_0 = h'(0)$ using
\begin{eqnarray*}
r_0  &=& -\dfrac{S'(0)^2 -S(0)S''(0)}{S(0)^2} 
\approx \left(\dfrac{S(\Delta t) - S(0)}{\Delta t S(\Delta t)}\right)^2  -\dfrac{S(2\Delta t) - 2 S(\Delta t) + S(0) }{\Delta t ^2 S(\Delta t)}.
\end{eqnarray*}
% \begin{eqnarray*}
% r_0  &=& -\dfrac{S'(0)^2 -S(0)S''(0)}{S(0)^2} 
% \approx \left(\dfrac{S'(\Delta t)}{S(\Delta t)}\right)^2  -\dfrac{S''(\Delta t)}{ S(\Delta t)} \\
% &\approx& \left(\dfrac{S(\Delta t) - S(0)}{\Delta t S(\Delta t)}\right)^2  -\dfrac{S(2\Delta t) - 2 S(\Delta t) + S(0) }{\Delta t ^2 S(\Delta t)}.
% \end{eqnarray*}
If there is uncertainty or reservations about these choices, one can choose a prior concentrated on such values. This could serve as an alternative method for estimating the initial conditions or for performing a sensitivity analysis \citep{christen:2024}.
In the example presented in Section~\ref{sec:example} the above method for setting the initial values $h_0$ and $r_0$ is used.

\section{Real data application}\label{sec:example}
%%%%%%%%%%%%%%%%%%%%%%%%%%%%%%%%%%%%%%%%%%%%%%%%%%%%%%%%%%%%%%%%%%%%%%%%%%%%%%%%%%%%%%%%%%%%%%%%%%%%%%
%%%%%%%%%%%%%%%%%%%%%%%%%%%%%%%%%%%%%%%%%%%%%%%%%%%%%%%%%%%%%%%%%%%%%%%%%%%%%%%%%%%%%%%%%%%%%%%%%%%%%%
In this Section, we present a real data application in which we analyze the \texttt{rotterdam} data set from the \texttt{survival} R package. This data set contains information about the survival times of $n=2,982$ breast cancer patients, of which $1,710$ cases were right-censored. Generally, breast cancer patients have a good prognosis, and the hazard function may begin to increase slowly from the time of diagnosis. Clinically, this indicates that the hazard function starts at a low point and grows slowly at the beginning.
Based on these points, and following the discussion presented in Section~\ref{sec:inference} to fix $h_0$ and $r_0$, we assume that $\Delta t = 1/12$ (one month), $S(0)=1$ (no deaths before the start of follow-up), $S(\Delta t) = 0.999$ and $S(2 \Delta t) = 0.998$ (\textit{i.e.}~one death per 1000 patients per month, immediately after the start of follow-up). These choices lead to the initial conditions $h_0=0.012$ and $r_0=0.00014$. Similar values for the initial conditions would be obtained with slight variations in the initial choices, provided they are consistent with the clinical context.

\begin{figure}[h!]
\centering
\begin{tabular}{c c}
\includegraphics[width=0.45\textwidth]{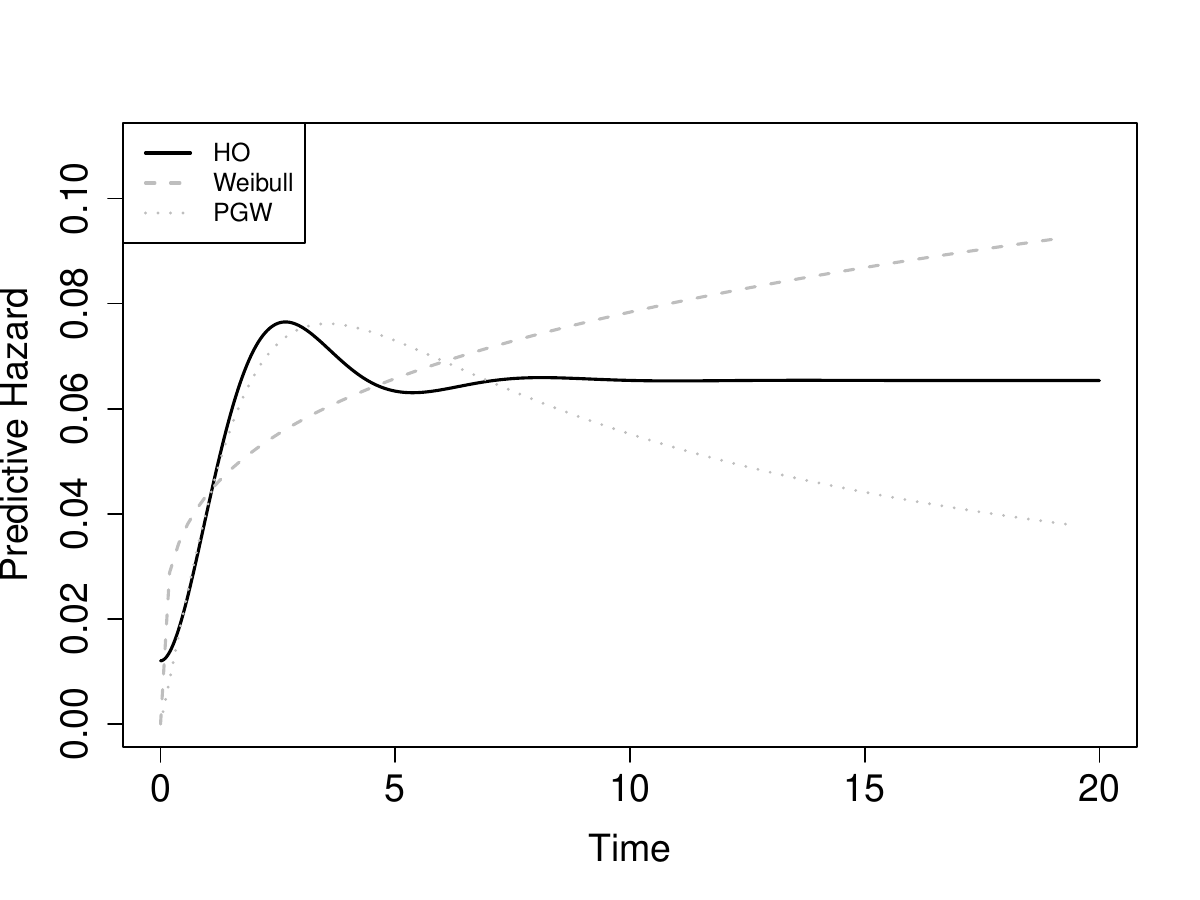} & \includegraphics[width=0.45\textwidth]{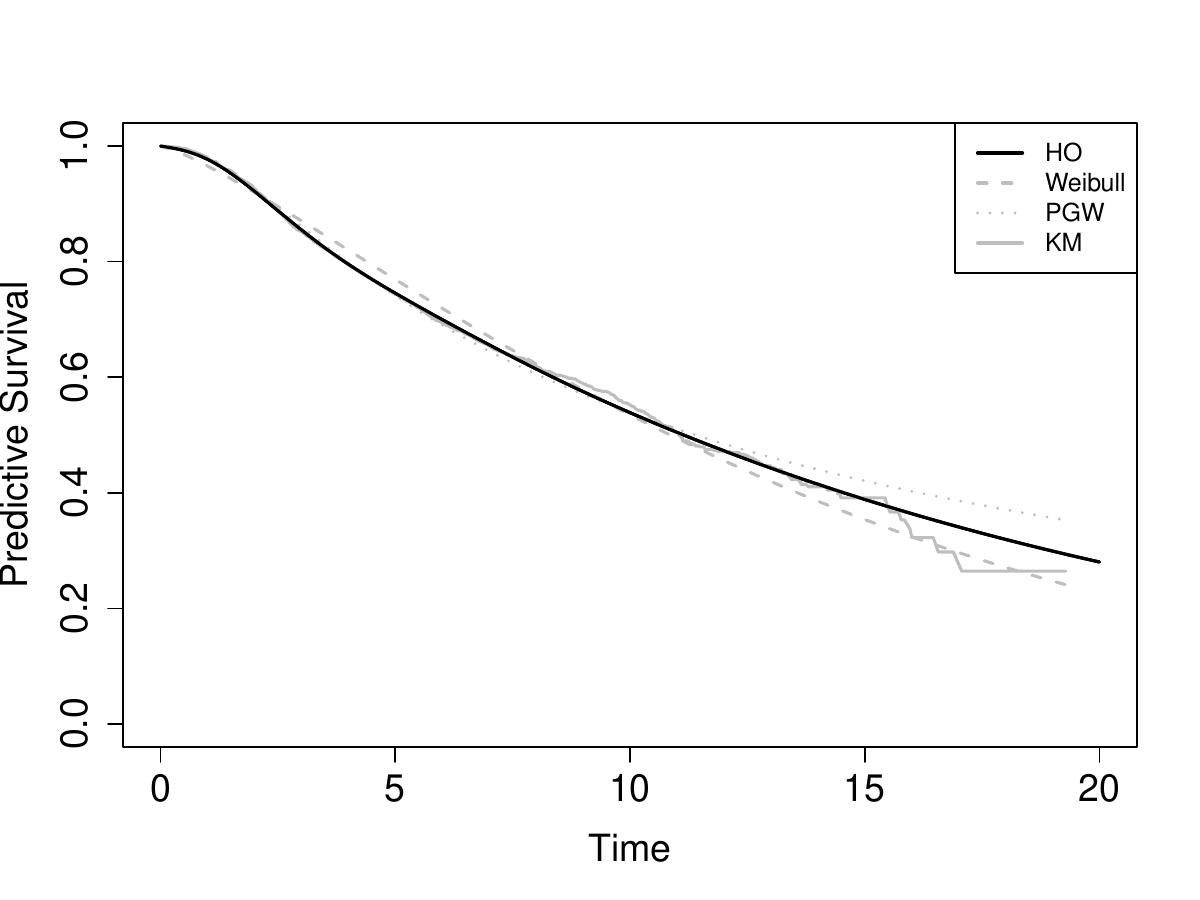} \\
 (a) & (b) \\
 \end{tabular}
\caption{\texttt{rotterdam} data. (a) Predictive hazard functions for the Weibull, PGW, and harmonic oscillator (HO) models; and (b) Predictive survival functions for the Weibull, PGW, and HO models, and Kaplan-Meier (KM) estimator.}
\label{fig:application}
\end{figure}

We fit the harmonic oscillator model \eqref{eq:harmonic_osc}, using a Bayesian approach with the fixed initial conditions mentioned above. The three parameters to be estimated are $\btheta = (\eta,w_0,h_b)$. We adopt a product prior structure
$$
\pi_{\btheta}(\btheta) \propto \pi( w_0) \pi(\eta) \pi(h_b), \quad \btheta \in \Theta_A,
$$
where $\pi$ is a gamma prior with a scale parameter $1000$ and shape parameter $0.001$. These are ``weakly informative'' priors which reflect a high degree of prior uncertainty about the parameters.
Figure \ref{fig:application} shows the predictive hazard and the predictive survival distributions for these three models, along with the Kaplan-Meier (KM) estimator of survival.

Finally, we compare the fit of the harmonic oscillator model \eqref{eq:harmonic_osc} against the Weibull distribution and the power generalized Weibull distribution (PGW), using the Bayesian information criterion (BIC). We choose weakly informative priors for the parameters of these models (gamma priors with scale parameter $1000$ and shape parameter $0.001$). The Weibull distribution can only capture increasing, decreasing or constant hazard shapes. The PGW distribution is a three-parameter distribution that can capture the basic shapes (except for the oscillatory shape). The BIC for the Weibull model is $9,650.30$, the BIC for the PGW model is $9,590.03$, and the BIC for the harmonic oscillator model is $9,581.037$. Thus, the harmonic oscillator model is clearly favored by the data, based on the BIC.

% %%%%%%%%%%%%%%%%%%%%%%%%%%%%%%%%%%%%%%%%%%%%%%%%%%%%%%%%%%%%%%%%%%%%%%%%%%%%%%%%%%%%%%%%%%%%%%%%%%%%%%
% %%%%%%%%%%%%%%%%%%%%%%%%%%%%%%%%%%%%%%%%%%%%%%%%%%%%%%%%%%%%%%%%%%%%%%%%%%%%%%%%%%%%%%%%%%%%%%%%%%%%%%
\section{Discussion}
% %%%%%%%%%%%%%%%%%%%%%%%%%%%%%%%%%%%%%%%%%%%%%%%%%%%%%%%%%%%%%%%%%%%%%%%%%%%%%%%%%%%%%%%%%%%%%%%%%%%%%%
% %%%%%%%%%%%%%%%%%%%%%%%%%%%%%%%%%%%%%%%%%%%%%%%%%%%%%%%%%%%%%%%%%%%%%%%%%%%%%%%%%%%%%%%%%%%%%%%%%%%%%%

We have developed a novel parametric hazard model obtained by enforcing positivity in the damped harmonic oscillator. As shown in the qualitative analysis of the solutions to the corresponding second order ODE, the parameters of this model are interpretable and the model is both tractable and flexible. The real data analysis presented here shows that the proposed model offers competitive performance compared to flexible parametric models commonly used in survival analysis. 

{The focus of this paper has been on survival analysis without covariates. A potential extension involves incorporating covariates into the damped harmonic oscillator hazard function proposed here. This can be achieved by using this model as the baseline hazard in any hazard-based regression model \citep{rubio:2019}, or by introducing covariates through linear predictors on the parameters (similar to distributional regression). These directions will be explored in future work.}

%While we provided a strategy for constructing priors in the under-damped case in Section \ref{sec:example}, developing informative priors for the over-damped case, as well as objective priors for all regimes, represents a potential direction for future research.

\bibliographystyle{plainnat}
\bibliography{references}

\clearpage
\appendix
%%%%%%%%%%%%%%%%%%%%%%%%%%%%%%%%%%%%%%%%%%%%%%%%%%%%%%%%%%%%%%%%%%%%%%%%%%%%%%%%%%%%%%%%%%%%%%%%%%%%%%
%%%%%%%%%%%%%%%%%%%%%%%%%%%%%%%%%%%%%%%%%%%%%%%%%%%%%%%%%%%%%%%%%%%%%%%%%%%%%%%%%%%%%%%%%%%%%%%%%%%%%%
\section*{Appendix}\label{sec:appendix}
%%%%%%%%%%%%%%%%%%%%%%%%%%%%%%%%%%%%%%%%%%%%%%%%%%%%%%%%%%%%%%%%%%%%%%%%%%%%%%%%%%%%%%%%%%%%%%%%%%%%%%
%%%%%%%%%%%%%%%%%%%%%%%%%%%%%%%%%%%%%%%%%%%%%%%%%%%%%%%%%%%%%%%%%%%%%%%%%%%%%%%%%%%%%%%%%%%%%%%%%%%%%%

%%%%%%%%%%%%%%%%%%%%%%%%%%%%%%%%%%%%%%%%%%%%%%%%%%%%%%%%%%%%%%%%%%%%%%%%%%%%%%%%%%%%%%%%%%
\subsection*{Proof of Proposition 1}
%%%%%%%%%%%%%%%%%%%%%%%%%%%%%%%%%%%%%%%%%%%%%%%%%%%%%%%%%%%%%%%%%%%%%%%%%%%%%%%%%%%%%%%%%%
The proof follows by conducting a qualitative analysis of the solutions, along with classical results on the damped harmonic oscillator \citep{georgi:1993,strogatz:2018}.
\begin{itemize}
\item For $\eta > 1$, the system is in the over-damped case. $r_0 > 0$ implies that the hazard function is increasing at $t=0$, and $\frac{(h_0-h_b -a)}{(h_0-h_b +a)(w_1-w_0\eta)} < 0$ implies that the hazard function has no critical or inflection point. Consequently, the hazard function is monotonically increasing. 

\item For $\eta > 1$, the system is in the over-damped case. $r_0 < 0$ implies that the hazard function is decreasing at $t=0$, and $\frac{(h_0-h_b -a)}{(h_0-h_b +a)(w_1-w_0\eta)} < 0$ implies that the hazard function has no critical or inflection point. Consequently, the hazard function is monotonically decreasing. 

\item For $\eta > 1$, the system is in the over-damped case. $r_0 > 0$ implies that the hazard function is increasing at $t=0$, and $\frac{(h_0-h_b -a)}{(h_0-h_b +a)(w_1-w_0\eta)} > 0$ implies that the hazard function has a critical or inflection point. Consequently, the hazard function has initially an increasing behavior that changes to decreasing after the inflection point. This implies that the hazard is unimodal (up-then-down).

\item For $\eta > 1$, the system is in the over-damped case. $r_0 < 0$ implies that the hazard function is decreasing at $t=0$, and $\frac{(h_0-h_b -a)}{(h_0-h_b +a)(w_1-w_0\eta)} > 0$ implies that the hazard function has a critical or inflection point. Consequently, the hazard function has initially an decreasing behavior that changes to increasing after the inflection point. This implies that the hazard is bathtub shaped (down-then-up).

\item If $\eta<1$, then the system is in the under-damped case, which implies that the hazard function exhibits an oscillatory behaviour that stabilises due to damping \citep{georgi:1993}. 

\end{itemize}

%%%%%%%%%%%%%%%%%%%%%%%%%%%%%%%%%%%%%%%%%%%%%%%%%%%%%%%%%%%%%%%%%%%%%%%%%%%%%%%%%%%%%%%%%%
\subsection*{Proof of Proposition 2}
%%%%%%%%%%%%%%%%%%%%%%%%%%%%%%%%%%%%%%%%%%%%%%%%%%%%%%%%%%%%%%%%%%%%%%%%%%%%%%%%%%%%%%%%%%

% For the damped harmonic oscillator as in \eqref{eq:harmonic_osc} it is known that the real part of the roots of the characteristic polynomial are always negative \citep{georgi:1993,strogatz:2018} and therefore $h(t \mid \btheta) - h_b = \bo(e^{-\alpha t})$ for some $\alpha > 0$. That is, asymptotically, in the over--, under-- or critically-- damped cases, $h(t \mid \btheta)$ tends to $h_b$ exponentially fast as $t\to\infty$. This result, together with the definition ${\mathbb P}(T> t \mid \btheta) = \exp\left\{-\int_{0}^{t} h(r \mid \btheta) dr\right\}$, implies that there exists $M>0$ such that ${\mathbb P}(T> t \mid \btheta) \leq M \exp\left\{- h_b t\right\}$, from which the result follows. 

\begin{itemize}
\item[(i)] In the under-damped case ($\eta < 1$), we have that
\begin{eqnarray*}
{\mathbb P}(T>t \mid \btheta) &=& \exp\left\{ -H(t \mid \btheta)\right\} \\
&=& \exp\left\{ - h_b t \right\} \exp\left\{ - \frac{A}{w_0 \eta} 
\frac{\sin(\phi) + \mu \cos(\phi) }{\mu^2 + 1}  \right\}  \\
&\times& \exp\left\{\frac{A}{w_0 \eta}\frac{ e^{-  w_0 \eta t} (\sin(w_1 t + \phi) + \mu \cos(w_1 t + \phi))}{\mu^2 + 1}  \right\}.
\end{eqnarray*}
Now, using that $\frac{\sin(w_1 t + \phi) + \mu \cos(w_1 t + \phi)}{\mu^2 + 1}  \leq \frac{1 + \mu }{\mu^2 + 1} $, and denoting $K = \exp\left\{ - \frac{A}{w_0 \eta} \frac{\sin(\phi) + \mu \cos(\phi) }{\mu^2 + 1}  \right\}$, $K_1 = \dfrac{A}{w_0 \eta}  \dfrac{1+\mu}{1+\mu^2}$, and $K_2 = w_0 \eta$, it follows that
\begin{eqnarray*}
{\mathbb P}(T>t \mid \btheta) \leq K  e^{- h_b t + K_1 e^{-K_2 t}}.
\end{eqnarray*}
Therefore, we have shown that ${\mathbb P}(T> t \mid \btheta)  = \bo \left(  e^{- h_b t + K_1 e^{-K_2 t}  } \right)$.

\item[(ii)] In the over-damped case ($\eta > 1$), we have that
\begin{eqnarray*}
{\mathbb P}(T> t \mid \btheta)  &=& \exp\left\{ -H(t \mid \btheta)\right\} \\
&=& \exp\left\{ - h_b t \right\} \exp\left\{ - \frac{h_0-h_b + a}{2 (-w_0 \eta + w_1)}  (e^{ (-w_0 \eta + w_1) t} - 1)\right\} \\
&\times&  \exp\left\{ - \frac{h_0-h_b - a}{2 (-w_0 \eta - w_1)} (e^{(-w_0 \eta - w_1) t} -1) \right\}.
\end{eqnarray*}
Now, denoting $K =  \exp\left\{  \frac{h_0-h_b + a}{2 (-w_0 \eta + w_1)} \right\} \exp\left\{ \frac{h_0-h_b - a}{2 (-w_0 \eta - w_1)}\right\}$,  $K_1 = \max\left\{ \dfrac{h_b-h_0-a}{w_1-w_0\eta}, \dfrac{h_0-h_b-a}{w_1+w_0\eta} \right\}$, and $K_2 = w_0 \eta -w_1$, it follows that
\begin{eqnarray*}
{\mathbb P}(T>t \mid \btheta) \leq K  e^{- h_b t + K_1 e^{-K_2 t}}.
\end{eqnarray*}
Therefore, we have shown that ${\mathbb P}(T> t \mid \btheta)  = \bo \left(  e^{- h_b t + K_1 e^{-K_2 t}  } \right)$.

\item[(iii)] In the critically-damped case ($\eta = 1$), we have that 
\begin{eqnarray*}
{\mathbb P}(T> t \mid \btheta)   &=& \exp\left\{ -H(t \mid \btheta)\right\} \\
&=& \exp\left\{ - h_b t \right\} \exp\left\{-\dfrac{(h_0-h_b) \left(1 - e^{- w_0 t } \right)}{  w_0} \right\}\\
&\times& \exp \left\{ - \dfrac{\left( r_0 + w_0 (h_0-h_b) \right)}{w_0^2 } \left( e^{w_0 t} - w_0 t - 1 \right) e^{-w_0 t} \right\}
\end{eqnarray*}

Now, denoting $K= \exp\left\{-\dfrac{(h_0-h_b) }{  w_0} \right\} \exp \left\{ - \dfrac{\left( r_0 + w_0 (h_0-h_b) \right)}{w_0^2 }  \right\}$, $K_1 = \max \left\{ \dfrac{h_0-h_b}{w_0}, \dfrac{r_0 + w_0(h_0-h_b)}{w_0^2} \right\}$, $K_2 = w_0$, and $K_3 = \dfrac{r_0 + w_0(h_0-h_b)}{w_0}$, it follows that
\begin{eqnarray*}
{\mathbb P}(T>t \mid \btheta) \leq K \exp\left\{ - h_b t \right\} e^{- h_b t + K_1 e^{-K_2 t}  + K_3 t e^{-K_2 t}}.
\end{eqnarray*}

Therefore, we have shown that ${\mathbb P}(T> t \mid \btheta)  = \bo \left(  e^{- h_b t + K_1 e^{-K_2 t}  + K_3 t e^{-K_2 t}  } \right)$.
\end{itemize}
The result follows by noting that $K_2>0$ in all cases, which implies that the argument of the exponential can be upper-bounded by $-h_b t + C$, for some constant $C$, and $t$ sufficiently large. Consequently, the survival function has exponential tails $\bo \left( e^{-h_b t}\right)$.

\end{document}